\documentclass[11pt]{article}
\mathchardef\mhyphen="2D 

\usepackage{lipsum} 

\usepackage{tipa} 
\usepackage{bm}
\pdfoutput=1
\usepackage{amsmath,amsfonts,amsthm}
\usepackage{esdiff}  
\usepackage{booktabs}  
\usepackage{url}  
\usepackage{cleveref}  
	\crefname{equation}{equation}{equations}
	\crefname{figure}{figure}{figures}	
	\crefname{table}{table}{tables}
\usepackage[skip=1.5pt,font=small]{caption}

\usepackage{caption}
\usepackage{bbm}
\usepackage{relsize}  
\usepackage[usenames,dvipsnames,svgnames,table]{xcolor}

\usepackage{graphicx}
\usepackage{hyperref}  
\usepackage{tikz}

\usepackage[sc]{mathpazo} 
\usepackage[T1]{fontenc} 
\usepackage{microtype} 

\usepackage{multicol} 
\usepackage[margin={1cm,1.5cm}]{geometry}
\usepackage{booktabs} 
\usepackage{float} 
\usepackage{hyperref} 

\usepackage{lettrine} 
\usepackage{paralist} 

\usepackage{abstract} 

\usepackage{titlesec} 
\renewcommand\thesection{\Roman{section}} 
\renewcommand\thesubsection{\Alph{subsection}} 
\titleformat{\section}[block]{\large\scshape\centering\bfseries}{\thesection.}{1em}{} 

\titleformat{\subsection}[block]{\scshape\centering}{\thesubsection.}{1em}{} 

\usepackage[export]{adjustbox}
\usepackage{fancyhdr} 
\DeclareCaptionFormat{myformat}{#1#2#3\hrulefill}
\usepackage{authblk}
\makeatletter
\renewcommand\AB@affilsepx{, \protect\Affilfont}
\makeatother
\title{\vspace{-15mm}\fontsize{16pt}{16pt}\selectfont\textbf{Colonel Mustard in the Aviary with the Candlestick\\ \textit{a limit cycle attractor transitions to a stable focus via supercritical Andronov-Hopf bifurcation}}} %
\author[1]{Eve Armstrong\thanks{earmstrong@sas.upenn.edu}}
\affil[1]{Computational Neuroscience Initiative, University of Pennsylvania, Philadelphia, PA 19104} 

\date{(Dated: April 1, 2018)\vspace{-2ex}}
\renewcommand\Affilfont{\itshape\small}

\begin{document}
\maketitle 
\begin{abstract}
We establish the means by which Mr. Boddy came to transition from a stable trajectory within the global phase space of Philadelphia, Pennsylvania to a stable point on the cement floor of an aviary near the west bank of the Schuylkill River.  There exist no documented murder motives, and so the dynamical interaction leading to the crime must be reconstructed from circumstantial data.  Our investigation proceeds in two stages.  First we take an audio stream recorded within the aviary near the time of death to identify the local embedding dimension, thereby enumerating the suspects.  Second, we characterize Mr. Boddy's pre- and post-mortem behavior in the phase space in terms of an attractor that undergoes an abrupt change in stability.  A supercritical Andronov-Hopf bifurcation can explain this transition.  Then we uniquely identify the murderer.  Finally, we note long-term plans to construct an underlying dynamical model capable of predicting the stability of equilibria in different parameter regimes, in the event that Mr. Boddy is ever murdered again.
\end{abstract}
\hspace{0.8mm}
\begin{multicols}{2}
\section{INTRODUCTION}
Mr. Boddy Sylvester van Meersbergen was found lying unresponsive on the concrete floor of an aviary in Philadelphia, Pennsylvania in the pre-dawn of Sunday, April 1, 2018, in a congealed pool of his own blood (PPD Homicide, 2018).  The blood had emanated from Mr. Boddy's skull, which had suffered impact delivered by a dense blunt instrument.  Within the blood lay enmeshed twelve hulled sesame seeds, which are believed to have spilled from an unsealed zip-lock pouch marked \lq\lq For the Birds\rq\rq, lying several inches from Mr. Boddy's outstretched left hand.  

The aviary (Figure 1) is the property of the University of Pennsylvania.  It represents a collaboration between the Department of Neurocanticumology and the ambitious new multidisciplinary Initiative for the Understanding of Understanding.  The collaboration examines the role of song vocalization in avian social dynamics, and the neural basis for that song generation.  To that end, the aviary is wired with microphones.  These microphones record sound 24 hours per day.  
\begin{figure}[H]
\centering
  \includegraphics[width=60mm]{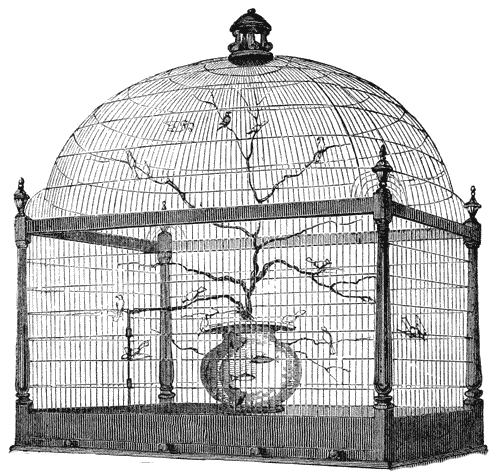}
  \caption{\textbf{Assumed phase-space location of the bifurcation.}}
\end{figure}
On the morning in question, the inanimate objects in the aviary were as follows (Figure 2): a plastic water dish, a plastic food dish, two stray zip ties, a roll of duct tape, a candlestick employed as a doorstop, and a feather duster.  In addition, the aviary, which is designed to attract local birds, currently contains roughly 40 birds of various species.  Most are shiny cowbirds (\textit{Molothrus bonariensis}), and occasionally a house sparrow (\textit{Passer domesticus}) will fly in.
\begin{figure}[H]
\centering
  \includegraphics[width=60mm]{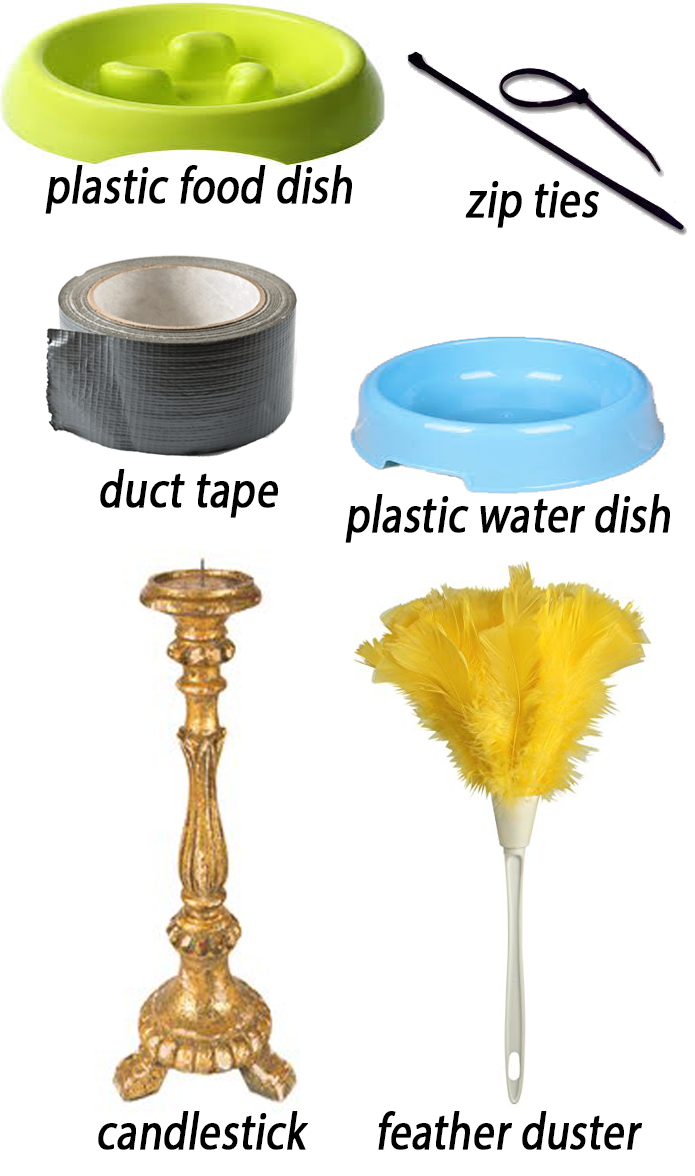}
  \caption{\textbf{Possible instruments of bifurcation via a dense blunt instrument.}}
\end{figure}
Mr. Boddy had no official affiliation with this aviary.  Rather, he was a former employee of a nearby Cheesecake Factory that recently had been condemned.  He had taken to whiling away his days of unemployment coverage meandering in a cycle about the neighborhood.  This cycle, which repeated reliably, consisted of the following locations: the aviary, a rocky outlook along the Schuylkill riverbank, a cherry tree, and a different cherry tree.

We use circumstancial evidence to identify the location and instrument of death: aviary and candlestick, respectively (see \textit{Methods}).  Identifying the murderer, however, is non-trivial.  While the Deceased's associations with the aviary inhabitants are documented, it seems that no murder motives exist (Cluedo, 1950).  That is: there exists no model relating variables of the system in a manner that identifies which, or even how many, of these birds should be considered suspects.  Moreover, there exists no represention of the underlying dynamics of this aviary space, nor of the space surrounding it.

We do, however, know something about the space: the stabilities of two equilibrium states within it.  As stated, prior to death Mr. Boddy reliably defined a route among four local neighborhoods.  Following death, Mr. Boddy reliably was dead.  There exists an apt framework for analyzing how an otherwise-stable equilibrium can instantaneously transition to another: the dynamical systems approach - and the construct of an attractor.

An attractor is a point, or a set of points, in a phase space to which all nearby trajectories will tend to converge.  This \lq\lq nearby\rq\rq\ region is the attractor's basin of attraction, or its global neighborhood.  An attractor may consist, for example, of a series of repeating locations in time (a limit cycle).  Such an attractor may represent a beating heart (Goldberger 1991), periodic electrical activity of neurons in the brain (Izhikevich 2007), gum chewing (Thelen 1989), or Mr. Boddy's regular trajectory prior to the murder.  Alternatively, an attractor may consist of a single point, which handily captures Mr. Boddy's equation of state following the murder.  

Now, importantly: under manipulations of the underlying structure of its phase space, an attractor's stability can change.  In this investigation, we shall seek to identify the means by which a stable limit cycle can transition to a stable point.  One well-described means is the Andronov-Hopf bifurcation (Andronov 1973, and work by Henri Poincar{\'e} and Eberhard Hopf).  A bifurcation is a nearly-instantaneous transformation of a system's behavior, due to a smooth change of a particular parameter that governs its dynamics.  This murder investigation, then, becomes a quest for a particular source of system manipulation: one that is capable of finagling an Andronov-Hopf bifurcation.  

Before seeking the murderer, however, we must first enumerate the suspects.  That is: how large a dimensionality must we consider this aviary to have, in order to reconstruct the crime?  And here we arrive at the one tool we have at our disposal: the aviary's audio stream during the time window of death.  As we shall describe in \textit{Methods}, the dynamical systems framework permits the identification of the dimensionality inherent within a given data set.  In this way, we shall identify the number of individual agents who may have independently contributed to the crime.   

\textit{Methods}, then, proceeds in two stages: 1) dimensionality identification within the local neighborhood of the aviary in order to enumerate the suspects, and 2) stability analysis of the attractor, to identify the means of altering its global topology.  This second step requires an underlying dynamical model.  As stated, in this case we do not have one.  For future investigations, however, it will be instructive to develop one, and importantly one that predicts the effects upon attractor stability of tweaks to particular model parameters.  In the event that Mr. Boddy is ever murdered again, armed with such a roadmap we will be better prepared.

\section{METHODS}
\subsection{\textbf{The local neighborhood}}

First we identify the local neighborhood and instrument of death.  

While Mr. Boddy was discovered in the aviary, we would be remiss in overlooking three other locations within reasonable proximity.  All aviary supplies are stored in the kitchen of the previously-mentioned condemned Cheesecake Factory, which is an independent structure standing roughly one hundred paces from the aviary.  In addition, the aviary connects via tunnel to the first-floor conservatory in the Wharton School of Business, and to the Billiard Room \& Sports Bar in the Graduate Student Center lounge, both located on the main University of Pennsylvania campus\footnote{The motivation for this tunnel system has not been ascertained.}.  For simplicity we shall assume that the scene of death was the aviary.   

Second, we identify the instrument of death to be the candlestick, as it appears to be the only object in the local neighborhood capable of bludgeoning.

\subsection{\textbf{Enumeration of suspects within the local neighborhood}}

\subsubsection{Time-delay embedding}

Next we reconstruct the dimension of the audio data stream obtained from the local neighborhood, in order to identify the minimum number of suspects required to fully investigate the crime.  The six suspects depicted in Figure 3 were identified following the employment of this method.  

This construction in fact consists of two stages: identifying a \textit{global} dimension ($d_E$) in which the attractor lives, followed by a \textit{local} ($d_L$) dimension within the neighborhood of the aviary.  The latter dimension will correspond to the number of active degrees of freedom in that neighborhood.  The establishment of a global embedding di-
\end{multicols}
\begin{figure}[H]
\centering
  \includegraphics[width=\textwidth]{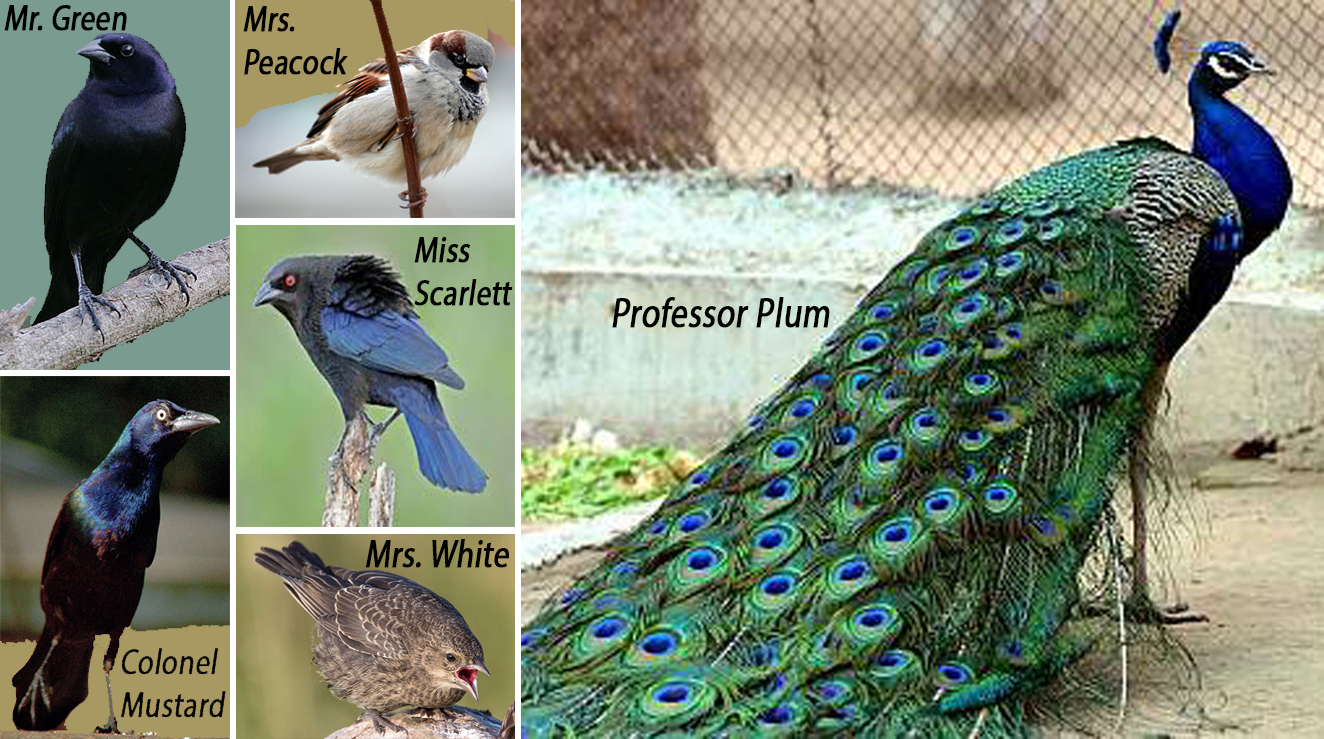}
  \caption{\textbf{The six suspects, where the enumeration of \lq\lq six\rq\rq\ was obtained via calculation of the local embedding dimension of the aviary.}  Clockwise from top left: a) \textit{Mr. Green}:  a cowbird.  Seems pretty normal.  b) \textit{Mrs. Peacock}: a sparrow who fell into the aviary and is too fat to escape.  c) \textit{Professor Plum}: a peacock.  Believed to be planted as a prank by one of the graduate students in Bio-engineering.  d) \textit{Mrs. White}: a cowbird.  e) \textit{Colonel Mustard}: a cowbird.  Likes to dance.  f) \textit{Miss Scarlett}: a cowbird; male.  Looks mean.}
\end{figure}
\begin{multicols}{2}
\noindent
mension $d_E$ is a critical first step because it defines for us a space that preserves all information - including the temporal structure - contained in the unknown dynamics that produced the audio stream\footnote{We note that because the audio stream was obtained within a local region of the global phase space, strictly speaking it cannot be used to extract a global dimension.  That is: we cannot assume that it represents the complete space.  In this case, however, we claim that this assumption is justified.  We have sampled Mr. Boddy's pre-mortem trajectory sufficiently to state with confidence that there is not all that much difference amongst the various local neighborhoods comprising Western Philadelphia.}.  

The method we will use to identify the global dimension present in a data set is based on the embedding theorem of Ma{\~n}{\'e} et al. (1981) and Takens (1981).  At its root is the following notion: one can represent a phase space in $n$ variables, or equivalently in one variable at $n$ distinct temporal locations (Ruelle 1979).  For example, one may use as independent coordinates $x$, $y$, and $z$, or just $x$ at times $t$, ($t - \tau$), and ($t - 2\tau$), where $\tau$ is a time delay between observations of $x$, chosen to render the time-delayed coordinates orthogonal.

Imagine that there exists some deterministic model $\bm{F}$ consisting of dynamical variables $\bm{birds}$.  $\bm{F}$ is a map telling us how to forward the state of $\bm{birds}$ in time from t $= n$ to ($n+1$).  In continuous time, this process is written as:
\begin{align}
  \diff{bird_{i}(t)}{t} &= f_i(\bm{birds}),
\end{align}
\noindent
where $bird_i$ is the $i^{th}$ bird and $f_i$ is the component of the map $\bm{F}$ dictating the evolution of $bird_i$.  $\bm{F}$ tells us how the birds interact with each other, and possibly with aspects of their environment.  

Now, imagine that we in fact do not have the map $\bm{F}$, and we would like to infer some of its basic properties.  Let's say that we do have a set of scalar measurements $data(\tikz\draw[black,fill=black] (0,0) circle (.5ex);)$ of some vector function $\bm{g}(\bm{birds}(n))$, which presumably were generated by $\bm{F}$.  If we have $data$, then we can \lq\lq unfold\rq\rq\ the geometric structure of the unknown dynamics $\bm{F}$ that produced it, in a new space constructed of new vectors.  These new vectors $\bm{y}(n)$ are:
\begin{align*}
  \bm{y} &= [data(\bm{birds}(n)),data(\bm{g}^{T_1}(\bm{birds}(n)),\\
  & data(\bm{g}^{T_2}(\bm{birds}(n)), \dots, data(\bm{g}^{T_{d-1}}(\bm{birds}(n)))].
\end{align*}
\indent
This is precisely the problem presented in the aviary.  We do not know the underlying dynamics $\bm{F}$ linking Mr. Boddy to any particular suspect $bird_i$, nor do we know which - or even how many - birds comprise the elements of the vector $\bm{birds}$.  We do, however, have the aviary's audio stream $audio$.  In this case, we can reverse-engineer some features of $\bm{F}$.  These features might turn out to be useful.

To proceed, let us make three simplifications.  First, $data(\bm{birds}(n))$ is simply the time series of our recorded aviary audio signal: $audio(n)$.  Second, the mapping $\bm{g}(\bm{birds}(n))$ takes $\bm{birds}$ at time $n$ to $\bm{birds}$ at one time delay $T_k$ later.  That is: the $T_k^{th}$ power of $\bm{g}$ is:
\begin{align*}
  \bm{g}^{T_k}(\bm{birds}(n)) = \bm{birds}(n + T_k).
\end{align*}
\noindent
Then our new vectors $\bm{y}(n)$ can be written as:
\begin{align*}
  \bm{y}(n) &= [audio(n), audio(n + T_1), \dots, audio(n + T_{d-1})].
\end{align*}
\noindent
Taking $T_k = kT$: 
\begin{align*}
  \bm{y}(n) &= [audio(n), audio(n + T), \dots, audio(n + (d-1)T)].
\end{align*}
\noindent
Third, we take $T$ to be an integer multiple of the sampling time $\tau_s$ of the audio signal.  So, the components of $\bm{y}$ are time lags $T\tau_s$ of $audio$.  The \textit{number of time lags} $d$ we call the \textit{dimension} of our newly-constructed phase space. 

\subsubsection{Choosing the time delay}

We seek to calculate $d$: the dimension of vectors $\bm{y}$.  To do this, we must first choose an appropriate time lag between the coordinates of $\bm{y}$.  This time lag should render the coordinates essentially independent of each other, so that they form an orthonormal basis that can completely define the phase space.

One way to require independence of coordinates $audio(n)$ and $audio(n + T\tau_s)$ is to plot their average mutual information\footnote{The average mutual information $AMI(audio(n)\mid audio(n + T\tau_s))$ is the amount of information, in bits, about the measurement $audio(n + T\tau_s)$ that is contained within the measurement $audio(n)$.  If $audio(n)$ and $audio(n + T\tau_s)$ are orthogonal, then neither should contain much of any information about the other.}:
\begin{align*}
  AMI(audio(n)\mid audio(n+T\tau_s))
\end{align*}
\noindent
as a function of $T$, and choose the value of $T$ that yields the first minimum of this distribution (Abarbanel 1996).  In this way, we calculated the optimal time delay given the data $audio$, and found that a value of 34 ms maximizes the independence among the coordinates\footnote{The time delay value of 34 ms may very well contain significance of its own; we haven't thought about it.}.

\subsubsection{Identifying the global dimension of the audio stream}

Armed with the optimal time delay, we are in a position to identify the number of time-delayed coordinates required to calculate the global embedding dimension contained in data $audio$.  Here we follow the approach taken by Kennel et al. (1992), which is explained in detail in Abarbanel (1996).  

Consider that the time series $audio(n)$ may be a projection of these data from a higher-dimensional space.  If this is the case, points in $audio(n)$ may appear nearby for one of two reasons: 1) the underlying dynamics render them nearby, in which case they are true \lq\lq nearest neighbors\rq\rq, or 2) the points appear to be nearby via the projection; these are false nearest neighbors.  We seek to remove the latter.

Say our vector $\bm{y}(n)$ has a nearest neighbor $\bm{y}^{NN}(n)$:
\begin{align*}
  \bm{y}(n) &= [audio(n), audio(n + T), audio(n + 2T), \\
  & \dots, audio(n + (d-1)T)];\\
  \bm{y}^{NN}(n) &= [audio^{NN}(n), audio^{NN}(n + T), audio^{NN}(n + 2T), \\
  & \dots, audio^{NN}(n + (d-1)T)].
\end{align*}
\noindent
Perhaps $\bm{y}^{NN}(n)$ is the vector truly just ahead or behind $\bm{y}(n)$ along the orbit.  Or perhaps it has been projected to appear that way.  To test, we add another time-delayed coordinate $audio(n + dT)$ to each vector $\bm{y}(n)$ and $\bm{y}^{NN}(n)$, and determine whether the Euclidean distance $\mid \bm{y}(n) - \bm{y}^{NN}(n)\mid$ between them increases.  If it does, then $\bm{y}^{NN}$ is a false nearest neighbor\footnote{More accurately, we require that the distance added by increasing the dimension by one not exceed the \lq\lq diameter\rq\rq\ of the attractor.  Or: if $\mid audio(n + dT) - audio^{NN}(n + dT)\mid / R_A$ is greater than a number of order two, then $\bm{y}^{NN}$ is a false nearest neighbor.  Here, $R_A$ - the radius of the attractor - is defined as the RMS value of the data $audio$ about its mean.}.

We iterate this process, each time adding another time-delayed coordinate to the length of each vector $\bm{y}$ and $\bm{y}^{NN}$, until we have removed all false nearest neighbors.  The number $d$ of coordinates we wind up with is the minimum dimensionality required to unfold the data.  This number corresponds to the global embedding dimension $d_E$.

\subsubsection{Identifying the local dimensionality of the audio stream}

Within the global phase space defined by $d_E$, it remains to calculate a \textit{local} dimension $d_L$ - or, the number of active degrees of freedom - in the vicinity of the aviary around the time of Mr. Boddy's death.  This step involves defining a field of \lq\lq neighboring\rq\rq vectors within this neighborhood, and requiring that the time evolution of these vectors be independent of $d_L$.  The explanation is rather tedious - even more tedious than that of the previous section, and we will spare you (see Abarbanel 1996).  

We calculated $d_L$ for the segment of the aviary's audio stream that spanned the five-hour time window of death that was inferred postmortem.  We identified a local dimension of 6.  Six independent agents contributed significantly to the dynamics within the aviary during this time.

\subsubsection{Identifying the six suspects}

It is tempting to seek to identify these six dimensions with six active agents, out of the $\sim$ 40 possibly-active agents within the aviary.  This can only be done in a rather ad-hoc way.  We, however, succumbed to temptation - and an ad-hoc way, due to the interesting coincidence that exactly six of the birds appeared particularly miffed\footnote{The miffedness was expressed vocally.} by our ongoing investigation at the crime scene.  The others did not take much notice; in fact, most slept through it.  Across the vast documentaries of make-believe murders, it is the general wisdom that the guilty conscience is the one unable to sleep (e.g. Shakespeare $\sim1600$, Poe 1843, Insomnia 2002).

The six suspects are shown in Figure 3.  Unless noted, they are cowbirds.  Clockwise from top left are: Mr. Green, Mrs. Peacock (a sparrow), Professor Plum (a peacock), Mrs. White, Colonel Mustard, and Miss Scarlett.  The peacock is believed to be a bit of tomfoolery played out by a graduate student in Bio-engineering.  

\subsection{\textbf{Behavior of the attractor in the global neighborhood}}

Let us return to the dynamical map $\bm{F}$.  Let us imagine that we know $\bm{F}$.  And now let us examine the attractors of such a network.  

Attractors are states of equilibrium, and their stability can be characterized.  To characterize an equilibrium, we linearize the functions $f_i$ (of Eq. 1) about each equilibrium, and construct a matrix of the partial derivatives of the $f_i$ with respect to the variables $birds_j$ evaluated at
\begin{figure}[H]
\centering
  \includegraphics[width=90mm]{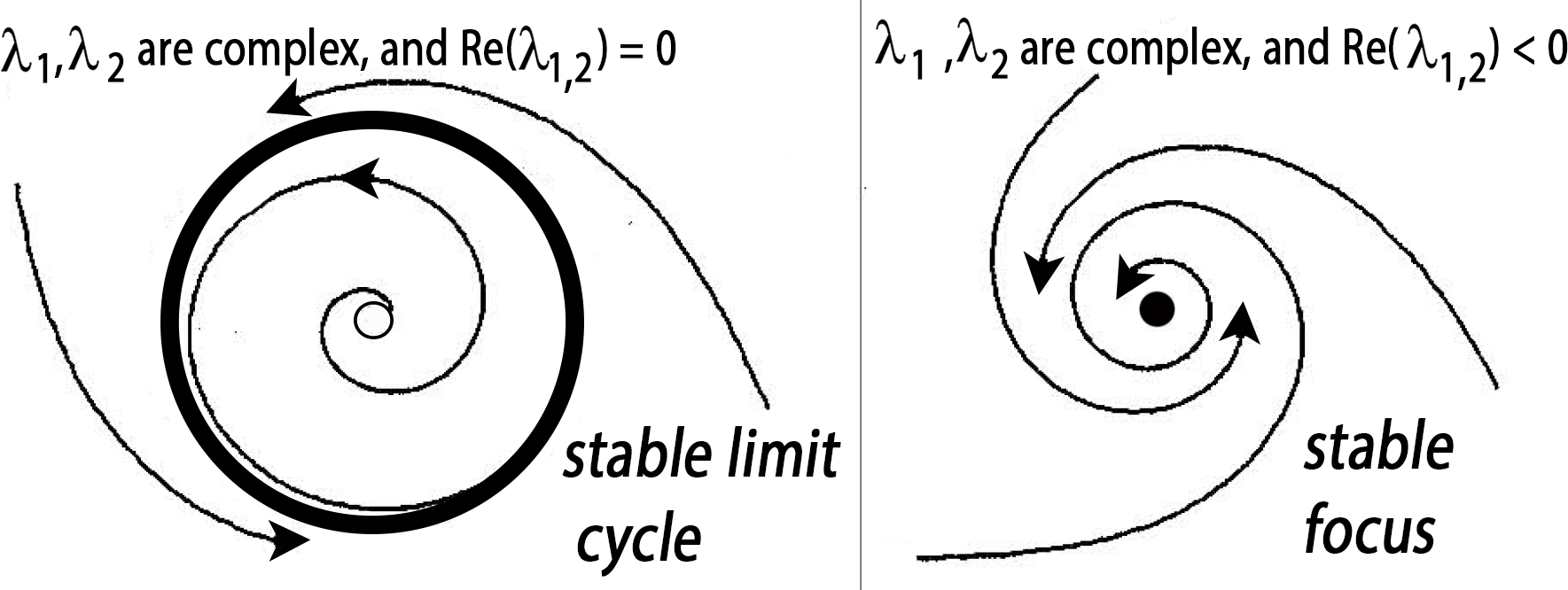}
  \caption{\textbf{Two types of attractor: a stable limit cycle} (left) \textbf{and stable focus} (right), represented in two dimensions.  In each case, all trajectories that begin within the neighborhood of the attractor converge to it.  A transition between the left and right panels represents an Andronov-Hopf bifurcation.}
\end{figure}
\noindent
that equilibrium.  This matrix is the Jacobian matrix $\bm{DF}$: 
\begin{align*}
\bm{DF} &= 
\begin{bmatrix}
\large{\frac{\partial f_1}{\partial bird_1}} & \dots & \frac{\partial f_1}{\partial bird_n} \\
\vdots & \ddots & \vdots \\
\frac{\partial f_{n}}{\partial bird_1} & \dots & \frac{\partial f_n}{\partial bird_n}.
\end{bmatrix}
\end{align*}
\indent
One way to determine the stability of a particular equilibrium is to examine the spectrum of eigenvalues and eigenvectors of $\bm{DF}$.  Numerical techniques can be found basically anywhere (von Mises \& Pollaczek-Geiringer 1929), and here we note the basics\footnote{We follow the QR algorithm (Francis 1961, Kublanovskaya 1961).}.  The eigenvalues $\lambda$ and corresponding eigenvectors $\bm{v}$ of $\bm{DF}$ are defined by the following relation: 
\begin{align*}
  DF \bm{v}_a &= \lambda_a \bm{v}_a.
\end{align*}  
\noindent
Each eigenvector $\bm{v}_a$ represents a direction in the phase space.  The eigenvalue $\lambda_a$ is a number associated with eigenvector $\bm{v}_a$ that dictates how trajectories along that direction $\bm{v}_a$ will behave near the equilibrium\footnote{Further, we can use the Jacobian matrix to examine how a phase space itself will behave in time.  To do so, we take the product of Jacobians along some trajectory from time t $= 0$ to $L$, denoted $\bm{DF}^L$, and examine the logarithms of the eigenvalues of the matrix $([\bm{DF}^L]^T \cdot \bm{DF}^L)^{1/{2L}}$.  In the limit $L \rightarrow \infty$, these are the global Lyapunov exponents of the system (Oseledec 1968).  They dictate the rates of growth or shrinking of volumes, for any volume of dimension equal to or less than the dimensionality of the global phase space $d$.  For a loose interpretation: if the global phase space is Philadelphia, then the values of the global Lyapunov exponents will dictate whether - and how quickly - Philadelphia will disappear, expand to Universal proportions, or continue as-is.}.  If the phase space is $d$-dimensional, then $a=[1,2,\dots,d]$.

Let's examine the relatively simple case of two dimensions, represented in Figure 4.  Here, we have two eigenvalue-eigenvector pairs.  If both eigenvalues of $\bm{DF}$ are real and negative, all nearby trajectories will converge to the (stable) equilibrium.  If both real parts are positive, then trajectories will diverge - the equilibrium is unstable.  If the eigenvalues contain imaginary parts, then nearby trajectories will contain a rotational component.  A transition from the left to right panel in Figure 4 represents an Andronov-Hopf bifurcation, which we will describe presently. 

Now let's examine Mr. Boddy's case, and treat it as two-dimensional.  Prior to death, his equilibrium can be characterized as a stable limit cycle attractor.  Such an equilibrium possesses purely imaginary eigenvalues: a trajectory neither grows nor shrinks, but rather rotates at a reliable rate.  Following death, Mr. Boddy's equilibrium is a single point.  What plausible route exists in a phase space for such a transition?

It is a general property of nonlinear systems that even a small change to a particular parameter within any of the functions $f_i$ can dramatically alter the landscape about an equilibrium of the system.  We call such a transition a bifurcation, and the particular parameter that effects the transition to be the \lq\lq bifurcation parameter\rq\rq.  Many types of bifurcations have been described in the literature; in this investigation we note a particularly pertinent one: the Andronov-Hopf bifurcation (see citations in \textit{Introduction}).  It is a well-described means to transition from stable limit cycle to stable point, where the point in this case is a focus: all nearby trajectories spiral toward it.  In two dimensions, this transition occurs when the real part of a complex-conjugate pair transitions from zero to negative.  

Upon retracing Mr. Boddy's precise orbit prior to death and his precise position within the aviary postmortem, we confirmed the legitimacy of this description\footnote{There are other possible explanations, but this one is sound and we're happy with it.}.

\subsection{\textbf{The murderer}}

It is an interesting enterprise to examine the role of a particular individual in the sculpting of a dynamical landscape.  To this end, it may seem tempting to seek an association between a particular eigenvalue/eigenvector pair and a particular individual.  This cannot be done\footnote{Yes, yet a second tempting thing that \lq\lq cannot be done\rq\rq.}.  What one can do, however, is to systematically alter the cast of individuals within any given landscape, and examine the resulting equilibria (or lack thereof) in each case.  This line of attack may reveal relationships among 
\end{multicols}
\begin{figure}[H]
\centering
  \includegraphics[width=120mm]{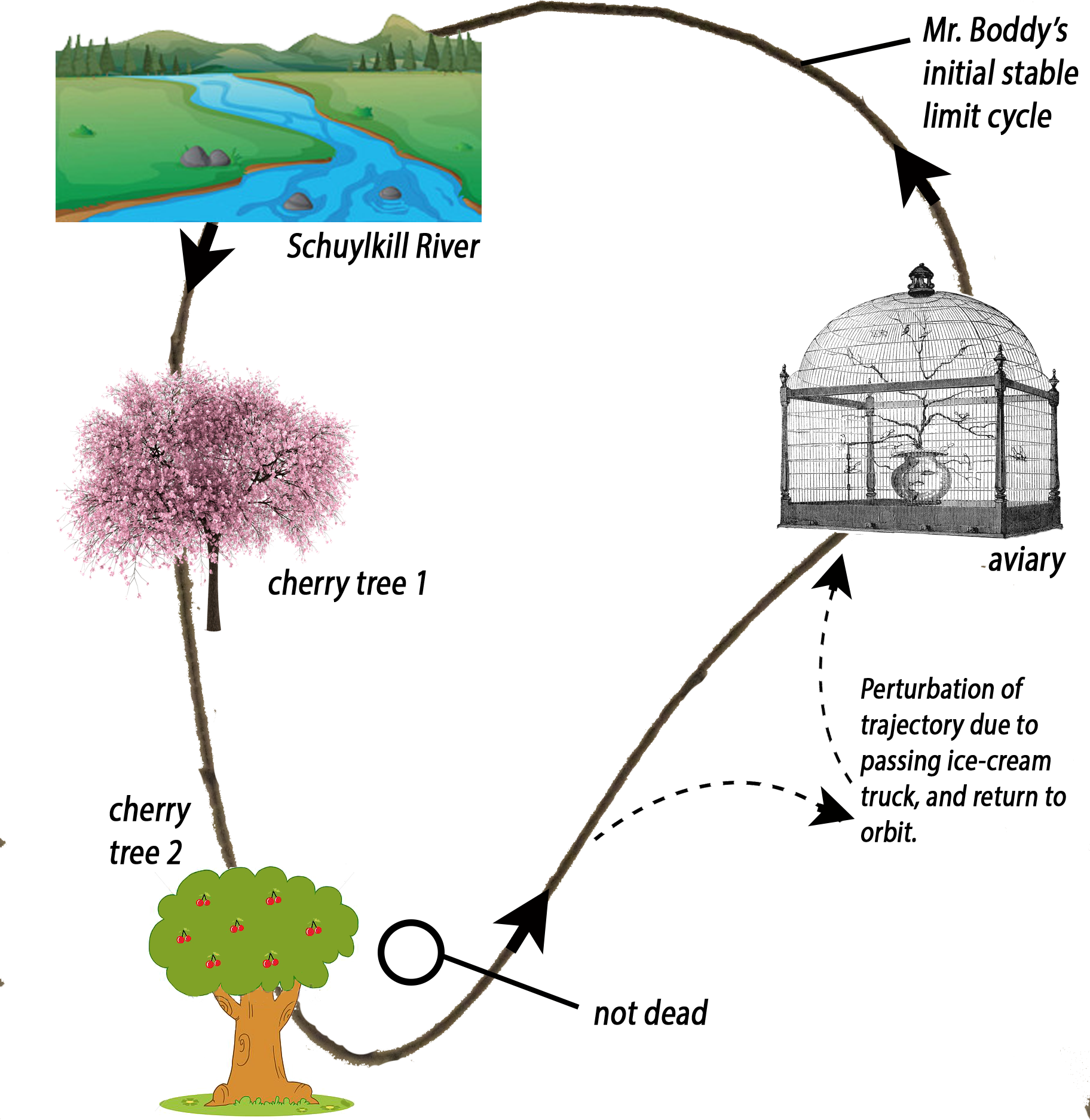}\vline
  \includegraphics[width=70mm]{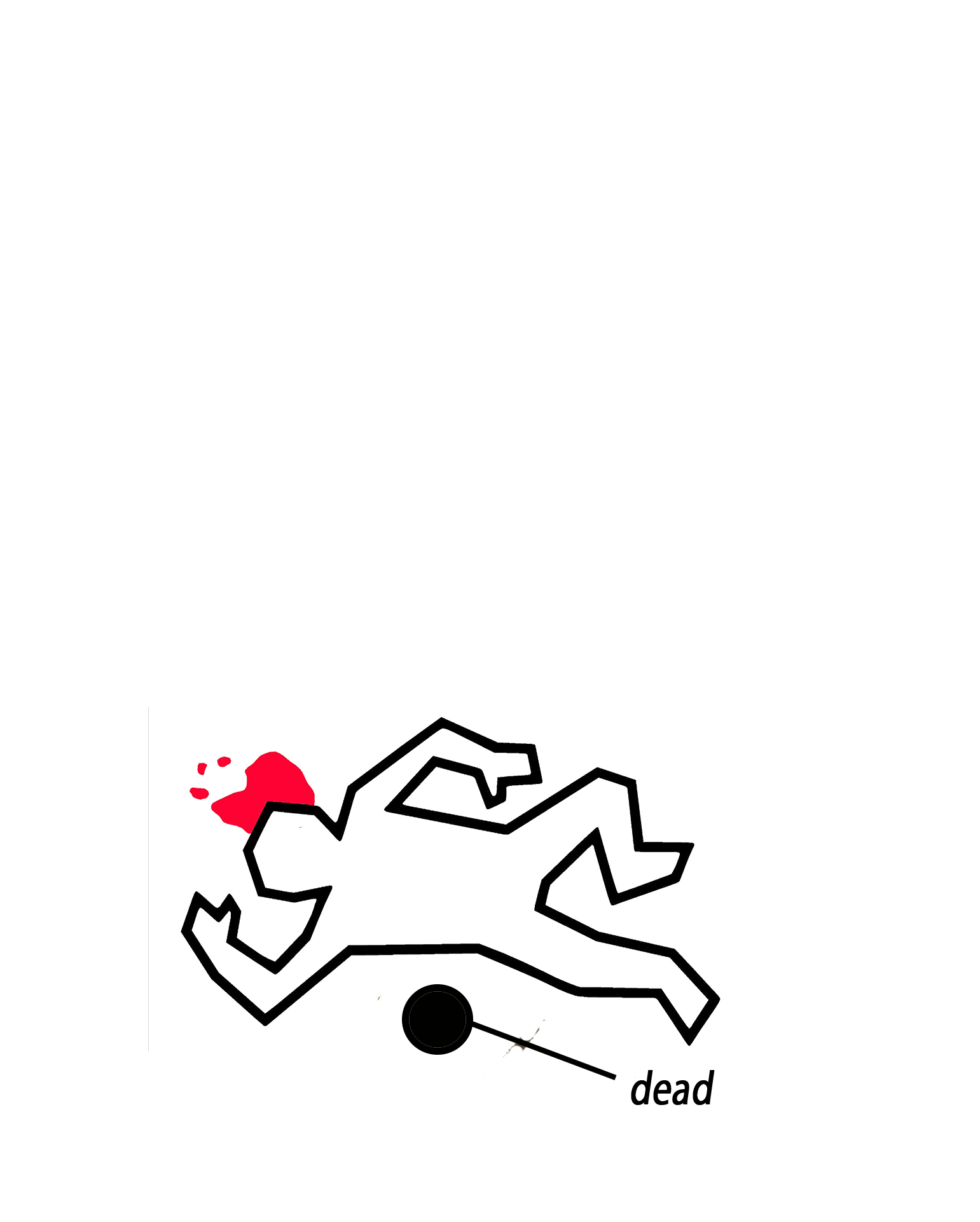}
  \caption{\textbf{The attractor framework applied to the phase space  of Mr. Boddy, prior to and following an Andronov-Hopf bifurcation.}  \textit{Left}: Before.  Mr. Boddy moves on a stable limit cycle attractor, amongst four local neighborhoods.  All trajectories in the neighborhood of this attractor converge to it.  For example, Mr. Boddy segues to a passing ice cream truck, but he returns to the orbit.  \textit{Right}: after.}
\end{figure}
\begin{multicols}{2}
\noindent
the chosen individuals, and eventually enable the reconstruction of the unknown dynamics $\bm{F}$ that map the landscape.  Then, working from $\bm{F}$ itself, one may identify bifurcation parameters with particular parameters in the dynamical equations, and investigate the stability of equilibria in different parameter regimes.  (For example: does altering a parameter in the equation $f_{m.scarlett}$ significantly alter the dynamical landscape in the neighborhood of the aviary?)  In principle, this approach could lead to a predictive model for criminal histories.

This approach is also beyond the scope of this investigation, which begs wrapping up.  For this reason, we shall dispense with further ruminations on the potential of dynamical systems for criminal reconstruction, and make a decision based on circumstance.  We settle on the suspect who appears at the moment to be the most suspicious: the one who flew the coop.  Colonel Mustard.  He finagled his way out through a tear in some chicken wire while we were distracted with the forensics team.  No one has seen him since.

\section{RESULT}
This investigation concludes that the Andronov-Hopf bifurcation of Mr. Boddy Sylvester van Meersbergen's equilibrium was pulled off by Colonel Mustard in the aviary with the candlestick\footnote{The means by which a 2.6-ounce bird wields a two-pound candlestick remains unclear.  We note, however, that this problem is not unprecedented.  The well-known swallow-vs-coconut problem (Monty Python, 1991) posed a similar question regarding weight ratios, and concluded that such a feat - while unlikely - may not be impossible.}.

\section{FUTURE WORK}
Several of UPenn's Medical School faculty - specifically a few in Otolaryngology - have recommended that in addition to examining the aviary's audio stream for the minimum dimensionality of its local phase space, that we also listen to it.  The motivation for this suggestion is to consider possible auditory cues such as gasps, moans, or muffled screams.  Such features might be taken as indicative of an ensuing murder, and might reveal likely means by which the act was conducted.  

Frankly, we are wary of human biases implicit in such a technique.  It is an interesting idea, however, for a complementary means of data analysis, and we plan to consider it in future work.
\end{multicols}
\newpage
\bibliographystyle{acm}
\nocite{*}
\bibliography{refs}

\end{document}